# Demonstration of Forward Inter-band Tunneling in GaN by Polarization Engineering

Sriram Krishnamoorthy, [1,a)] Pil Sung Park, [1] and Siddharth Rajan[1]

[1]Department of Electrical & Computer Engineering, The Ohio State University, Columbus, OH 43210

**Abstract**: We report on the design, fabrication, and characterization of GaN interband tunnel junction showing forward tunneling characteristics. We have achieved very high forward tunneling currents (153 mA/cm$^2$ at 10 mV, and 17.7 A/cm$^2$ peak current) in polarization-engineered GaN/InGaN/GaN heterojunction diodes grown by plasma assisted molecular beam epitaxy. We also report the observation of repeatable negative differential resistance in interband III-Nitride tunnel junctions, with peak-valley current ratio (PVCR) of 4 at room temperature. The forward current density achieved in this work meets the typical current drive requirements of a multi-junction solar cell.

---

[a)] Author to whom correspondence should be addressed. Electronic mail: krishnas@ece.osu.edu
Tel: +1-614-688-8458

Inter-band tunneling based devices have been investigated in a broad range of material systems since the discovery of quantum mechanical tunneling in degenerately doped Ge p-n junction reported by Esaki[1]. Extensive research has led to efficient tunnel junctions in material systems such as the III-As[2] and even indirect band gap material system such as the SiGe[3]. In the III-nitride system, efficient inter-band tunneling is inhibited by wider depletion regions due to the larger band gap and dopant solubility limitations in degenerately doped GaN p-n junctions. A solution to this problem is to use the high spontaneous and piezoelectric polarization charge[4,5] along the c-axis of III-nitrides for the design of tunnel junctions with greatly reduced tunneling barrier width[6,7,8,9]. The high polarization charge density at a polar hetero-interface creates a large band bending over a very small distance. With narrow band gap InGaN as barrier material, a very high current density of 118 A/cm$^2$ at a reverse bias of 1 V was been reported earlier by us [10]. However, the forward tunneling current density, required for applications such as such as multi-junction solar cells was still found to be low[10].

Forward tunneling in III- Nitrides is particularly interesting for photovoltaic application. The band gap of InGaN alloy spans the entire solar spectrum, but to exploit this band gap range, multi-junction solar cells with series connected InGaN active regions would be required. This necessitates efficient inter-band tunnel junctions with very low resistance under forward bias. In addition, p-type GaN has low hole diffusion length, and mobility and hence reduces solar cell efficiency. A thin p-type layer could be connected in series to a n-type layer through a tunnel junction to mitigate these effects.

In this work, we design and demonstrate high inter-band tunneling currents in GaN under forward bias using a GaN/InGaN/GaN heterostructure. We also report the observation of negative differential resistance in GaN based interband tunneling structure indicating efficient inter-band tunneling.

We first discuss the use of polarization engineering approach for design of a GaN/InGaN/GaN heterostructure tailored for forward tunneling. An In$_x$Ga1$_{-x}$N layer of thickness 't' sandwiched between GaN layers results in a band bending $\Phi(x)$ given by $\Phi(x) = \dfrac{q\,\sigma(x)\,t}{\varepsilon(x)}$, where $\varepsilon(x)$ is the permittivity of $In_xGa_{1-x}N$ and $\sigma(x)$ is the fixed polarization induced charge density at the $GaN/In_xGa_{1-x}N$ interface[11]. For a fixed composition of InGaN, as the thickness of the InGaN layer is increased, the space charge width of the n-GaN/InGaN/p-GaN junction reduces as the field due to the polarization charge dipole at the GaN/InGaN interface assists in dropping the built in potential of the junction. At a certain "critical" thickness, $t_{cr}$, the polarization dipole across the InGaN layer results in a band bending that equals the band gap of InGaN, as a result of which the conduction and valence bands are aligned on either side of InGaN. This is the optimum thickness of InGaN layer required to achieve very high tunneling current density in reverse bias as demonstrated earlier [10]. When thickness of the InGaN barrier layer is increased beyond $t_{cr}$, degenerate carrier gases start to accumulate at the GaN/InGaN interfaces. The equilibrium band diagram and the corresponding charge profile of a $GaN/In_{0.4}Ga_{0.6}N/GaN$ tunnel junction with 7 nm of InGaN, calculated using a self consistent Schrödinger poisson solver[12], is shown in Fig 1. At a low forward bias, there is resonant tunneling between the two dimensional degenerate carrier gases resulting in inter-band tunneling in GaN, as shown in the inset of Fig. 1. With further forward bias, the degenerate carrier gas density increases, but the conduction and the valence band edges are out of alignment, resulting in a sharp decrease in the tunneling current. This is manifested as the negative differential resistance regime. Beyond this, inter-band tunneling ceases and diffusion current across the GaN p-n junction becomes the dominant current. Thus, an InGaN barrier with thickness $t > t_{cr}$ can be utilized for enhancing the forward tunneling across degenerately doped GaN p- n junction, at the cost of reduced reverse tunneling current density owing to the increased thickness of InGaN layer beyond $t_{cr}$.

To demonstrate the use of GaN/InGaN/GaN tunnel junctions with high tunneling current density in forward bias, an InGaN TJ structure with $t > t_{cr}$ (Sample A) was grown on a Lumilog[13] N-polar free standing LED quality GaN template (dislocation density $\sim 10^8$ cm$^{-2}$) by plasma assisted molecular beam epitaxy in a Veeco Gen 930 system. The epitaxial structure of sample A is shown in the inset of Fig. 3. A reference sample with InGaN layer thickness $t < t_{cr}$ (Sample B), which is expected to show no forward tunneling characteristics, was grown for comparison. Sample B has a thinner quantum well (2 nm of $In_{0.3}Ga_{0.7}N$) than sample A, but in all other respects, is identical to sample A. The N-polar orientation of GaN was used in this work to achieve high composition InGaN for efficient inter-band tunneling. N-polar InGaN was grown using the conditions and growth model developed earlier[14,15]. InGaN was grown at a substrate temperature of $550^0$C followed by 70nm of p GaN ($N_A \sim 5 \times 10^{19}$ cm$^{-3}$). The structure was capped with 30 nm of highly Mg doped GaN cap layer to achieve ohmic contacts to the device. The thickness and composition of the InGaN layer was found to be 7 nm and 40 % respectively, from ω-2θ triple-axis scans using a high resolution X-ray diffractometer (not shown here). Ni/Au (20/150 nm) and Ti/Au (20/100 nm) stacks were evaporated for formation of ohmic contacts on p-GaN and n-GaN respectively.

The electrical characterization of these tunnel junctions is shown in Fig 2. Sample B with $t < t_{cr}$ exhibits orders of magnitude lower current density close to zero bias as compared to sample A as shown in Fig. 2 (c). Since the thickness of InGaN is not sufficient to drop the built in potential of the p-n junction, this sample shows poor tunneling characteristics in both forward and reverse bias. In contrast, sample A with $t > t_{cr}$ shows higher reverse current density due to tunneling, and also forward tunneling characteristics. A current density of 15 A/cm$^2$ is obtained at -1 V and there is a sharp increase in current density close to zero bias. However, this reverse current density is lower than the tunnel junction reported earlier, as this device is designed for forward tunneling characteristics at the expense of lowered reverse current density.

Two regimes are observed in the forward bias of the sample A shown in Fig 2 (a). At a very low forward bias, there is a sharp increase in the current density, which is in agreement with the design for forward tunneling. A current density of 153 mA/cm$^2$ is achieved even at a low forward bias of 10 mV compared to a drop of 1.3 V required in Sample B for the same current density. Hence this device would be an efficient tunnel junction for device applications requiring < 100 mA/cm$^2$ current density during forward bias of the tunnel junction. At higher forward bias, negative differential resistance regime is observed. The peak current density of 17.7 A/cm$^2$ is achieved at a forward bias of 0.8 V, followed by a sharp reduction in current density resulting in a peak- to valley current ratio (PVCR) of 4 at room temperature. The sharp reduction in current density is due to the step function nature of the two dimensional density of states. The higher peak current voltage can be attributed to the high ohmic contact resistance to p GaN. As the voltage is increased beyond the NDR onset voltage, the current is dominated by the diffusion current and excess current.

The device I-V was repeatable in both sweep directions (forward to reverse bias and vice versa) as long as the peak current voltage was not exceeded. When the device was biased beyond the NDR onset voltage, the device showed hysteresis in current with respect to sweep direction as shown in figure 3. During the reverse sweep (higher voltage to lower voltage sweep after a lower voltage to higher voltage sweep) NDR was not observed, and the current density was lowered. The current density was higher again at higher reverse bias and successive scans (low voltage to high voltage sweep) in forward direction shows NDR. We believe that the hysteresis observed here can be attributed to trapping effects related to a donor-like hole trap found at positive polarization charge interfaces close to the valence band edge that has also been found previously in N face HEMTs[16,17]. During the forward sweep, as the positive bias is increased, the net positive charge at the p-GaN/InGaN (top) interface increases due to increase in the density of holes and ionized donor-like traps. Eventually, the conduction and valence bands on either side cross, and the current drops. When the voltage is lowered again during the reverse sweep, the positive charge at the p-GaN/InGaN interface reduces due to recombination of holes or tunneling of the holes into the p-GaN. However, positively charged donor states *cannot* change their charge state. Due to this, the net electric field in the InGaN is *lowered* from its earlier value (before excess trap ionization), the entire energy band gap potential cannot drop across the InGaN layer, and an additional depletion layer is formed in the GaN. This additional depletion region increases the tunneling distance, thereby reducing the tunneling current greatly. However, under reverse bias, electrons from the p-GaN valence band may tunnel into the donor trap levels bringing the system back to its normal state, with clear negative differential resistance in the next forward sweep as observed in the experiment. We note that similar hysteresis with respect to sweep direction was observed in AlGaN/GaN double barrier resonant tunnel diodes[18].

Hysteresis occurs when the Fermi level moves below this trap and to avoid this, the device needs to be operated such that the Fermi level never moves below the trap level, which translates to device operating under low forward bias below the voltage at which the peak current is observed. In fact, as expected from these experiments, the tunnel junction *shows repeatable high current density* close to zero bias without any hysteresis, which is the regime in which a typical forward biased tunnel junction for a multi-junction solar cell is operated.

The demonstration of high tunneling current density at very low forward bias in this work, along with the earlier demonstration of high reverse tunneling current density shows the promise of polarization charge based approach for efficient inter-band tunneling structures, even in a wide band gap system such as III- Nitrides. With the appropriate designs, namely $t = t_{cr}$ and $t > t_{cr}$, one can achieve very high tunneling current densities in reverse bias and moderate tunneling current density in reverse bias with a very high current density in forward bias respectively. Since the tunneling probability goes exponentially with thickness of the barrier material, further optimization of the device structure is possible. We note that the tunnel junction demonstrated here would be absorbing the incoming solar radiation due to low band gap of InGaN barrier. However, the polarization engineering approach discussed here can also be utilized to design a GaN barrier between InGaN subcells.

In summary, we have demonstrated the use of polarization charge at the InGaN/GaN interface to design an inter-band tunneling structure with very high forward current density. The tunnel junction demonstrated here carries sufficient current density (150 mA/cm$^2$) at low forward bias (10 mV) to be incorporated in photovoltaic applications. We also report the observation of repeatable negative differential resistance at room temperature in any inter-band nitride tunneling structure, indicating efficient tunneling achieved in this wide band gap material system.

We would like to acknowledge funding from ONR under the DATE MURI program (Program manager: Paul Maki) .

**Figure captions**:

**Figure 1**: (Color online) (a) Equilibrium energy band diagram of n-GaN/In$_{0.4}$Ga$_{0.7}$N/p-GaN inter-band tunnel junction. **Inset**: Band diagram at low forward bias showing forward inter-band tunneling. (b) Charge profile showing the presence of degenerate carrier gases.

**Figure 2**: (Color online) Linear J-V characteristics of (a) Sample A with t > t$_{cr}$ designed for forward tunneling (b) Sample B with t < t$_{cr}$ showing no forward tunneling. (c) Comparison of log J-V characteristics of sample A and B.

**Figure 3**: (Color online) Log J-V characteristics of the GaN/In$_{0.4}$Ga$_{0.7}$N/GaN TJ showing hysteresis with respect to sweep direction. Negative differential resistance is reproducible with hysteresis. **Inset:** Epitaxial structure of the sample A.

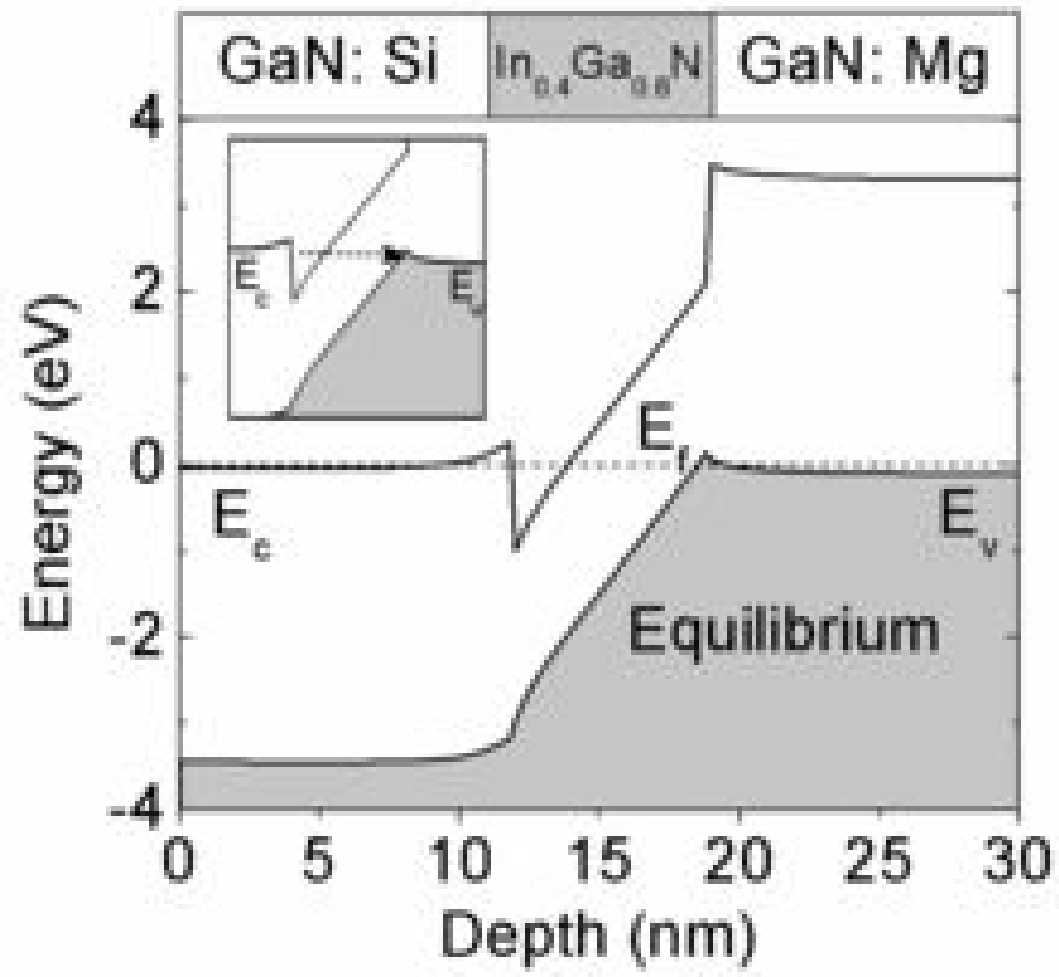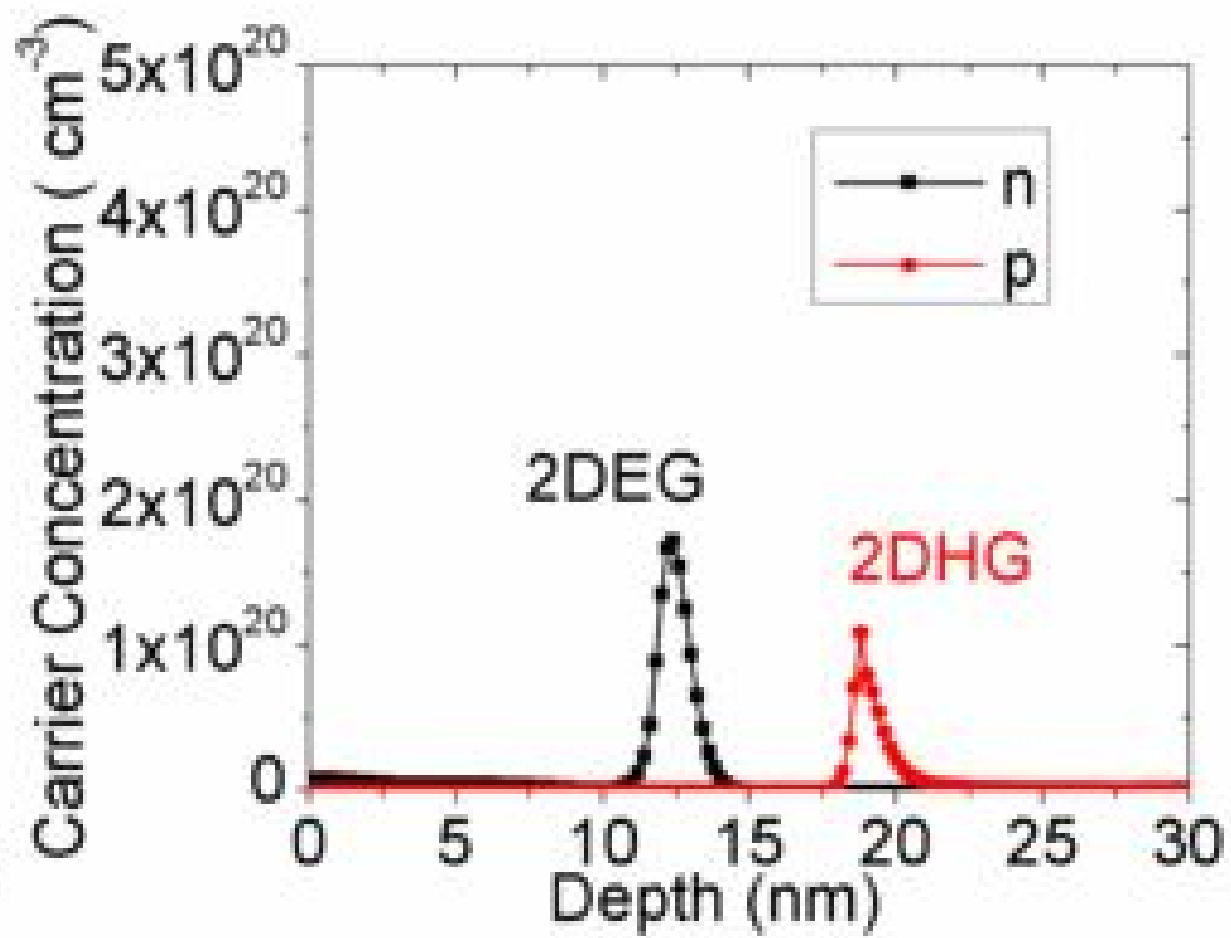

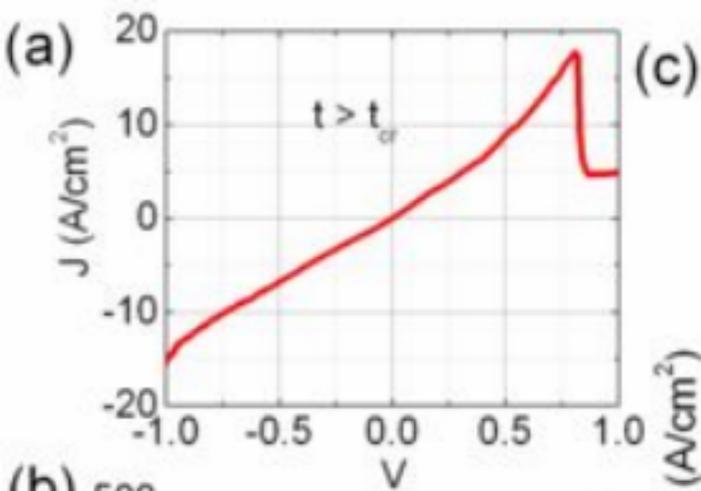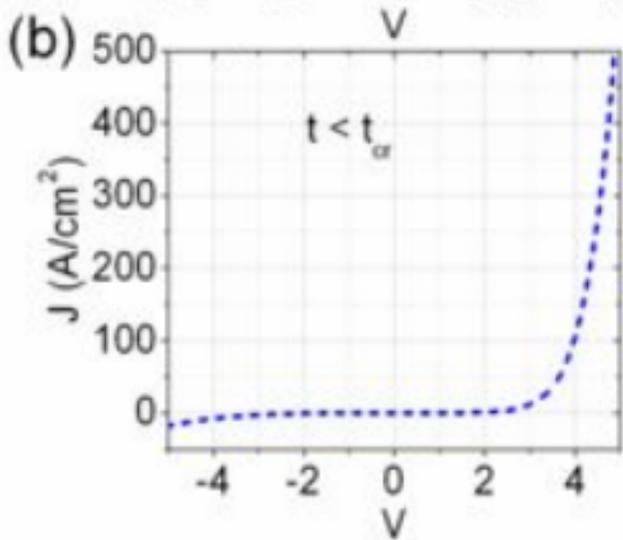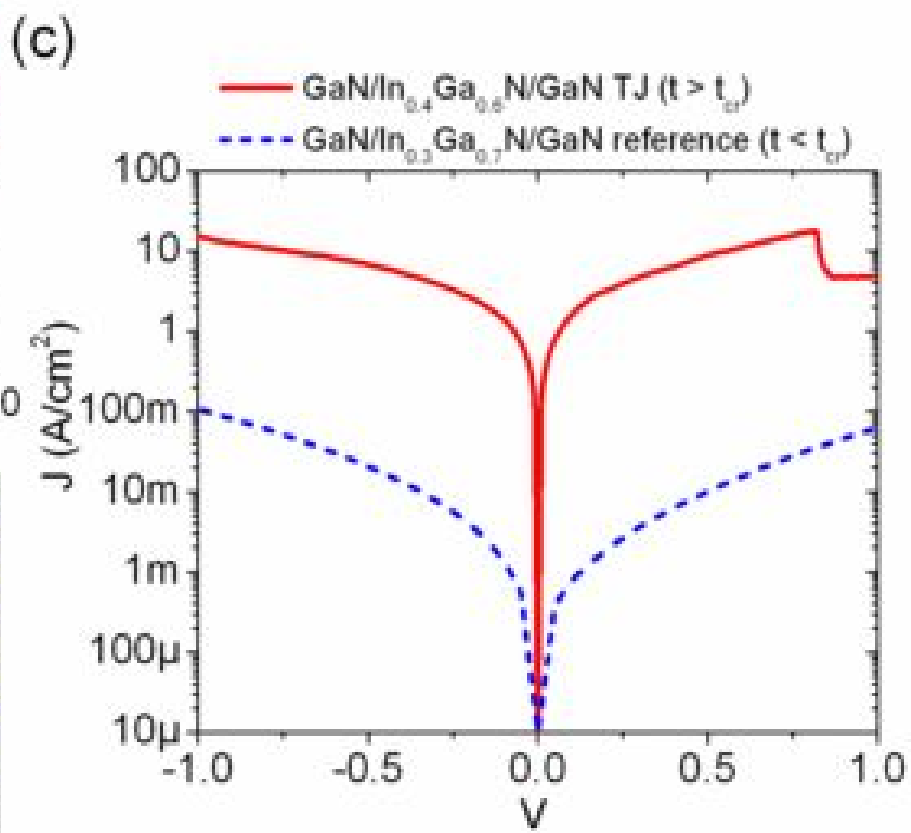

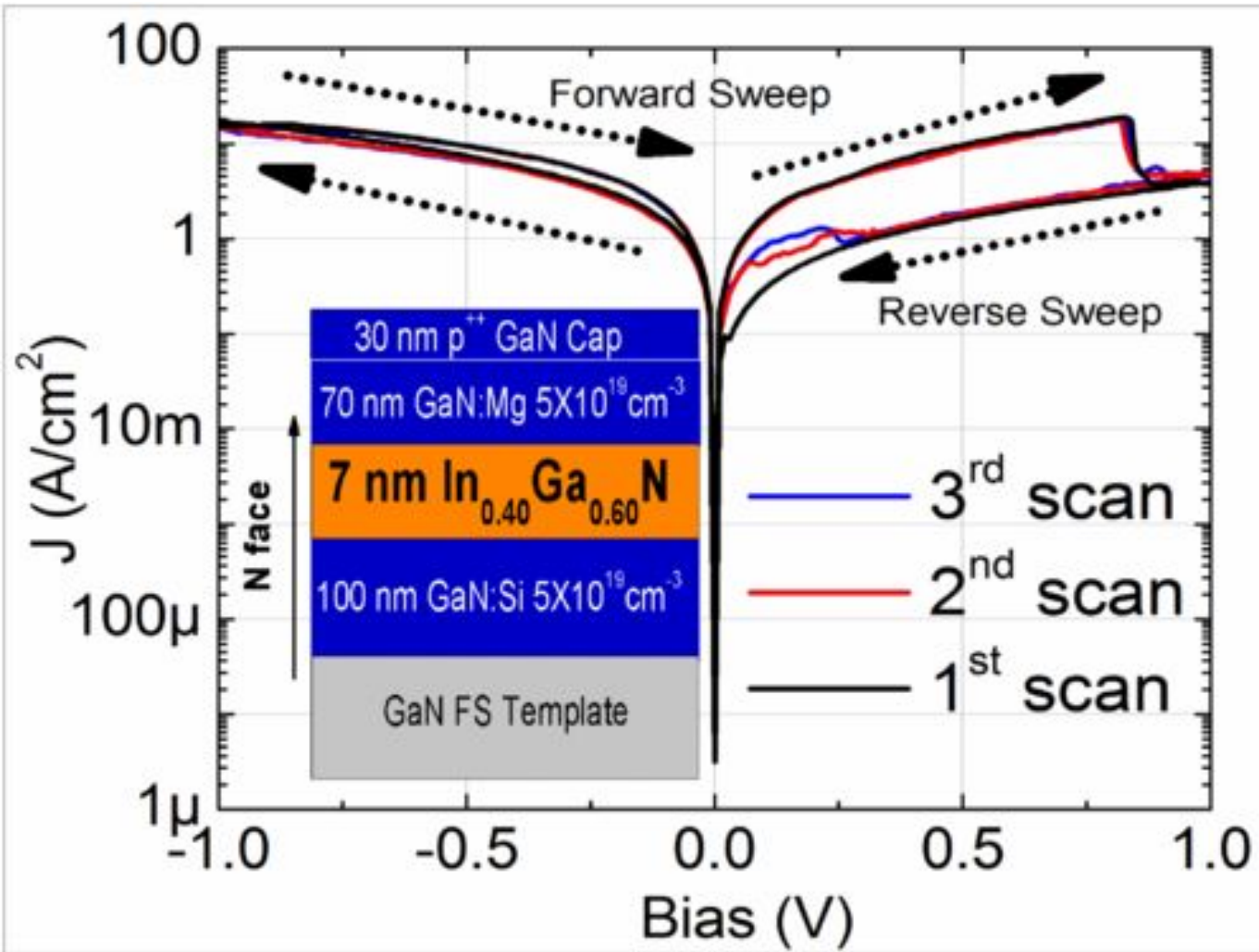